\documentclass[10pt, journal, letter, twocolumn]{IEEEtran} 
\usepackage{graphicx}
\usepackage{array}
\usepackage{multirow}
\usepackage{hhline}
\usepackage{amsthm}
\usepackage{epsfig}
\usepackage{latexsym}
\usepackage{amsfonts}
\usepackage{here}
\usepackage{rawfonts}
\usepackage[latin1]{inputenc}
\usepackage[T1]{fontenc}
\usepackage{calc}
\usepackage{capitalgreekitalic}
\usepackage{url}
\usepackage{enumerate}
\usepackage{color}
\usepackage[tbtags]{amsmath}
\usepackage{amssymb}
\usepackage{upref}
\usepackage{epic,eepic}
\usepackage{times}
\usepackage{dsfont}
\usepackage{comment}
\usepackage{cite}
\usepackage{mathrsfs}
\usepackage{verbatim}
\usepackage{float}
\usepackage{chngcntr}
\usepackage{bbm}
\usepackage{caption}
\usepackage{subcaption}
\allowdisplaybreaks
\usepackage{algorithm}
\usepackage{algpseudocode}

\newtheorem{proposition}{\bf Proposition}
\let\oldproposition\proposition
\renewcommand{\proposition}{\oldproposition\normalfont}

\newtheorem{definition}{\bf Definition}
\let\olddefinition\definition
\renewcommand{\definition}{\olddefinition\normalfont}

\newlength{\totlinewidth}
\newcounter{substep}

  {\end{list}}

\newlength{\aligntop}
\newlength{\alignbot}
 \makeatletter

\def\tablename{Table}

\IEEEoverridecommandlockouts

\usepackage{array}
\newcolumntype{P}[1]{>{\centering\arraybackslash}p{#1}}
\begin{document}
\title{\huge Integrated Millimeter Wave and Sub-6 GHz Wireless Networks: A Roadmap for Joint Mobile Broadband and Ultra-Reliable Low-Latency Communications}
%
\author{{Omid Semiari$^{1}$}, Walid Saad$^{2}$, Mehdi Bennis$^{3}$, and Merouane Debbah$^{4}$\vspace*{0em}\\
\authorblockA{\small $^{1}$Department of Electrical and Computer Engineering, Georgia Southern University, Statesboro, GA, Email: \protect\url{osemiari@georgiasouthern.edu}\\
	 $^{2}$Wireless@VT, Bradley Department of Electrical and Computer Engineering, Virginia Tech, Blacksburg, VA, USA, Email: \protect\url{walids@vt.edu} \\
$^{3}$Center for Wireless Communications, University of Oulu, Finland, Email: \protect\url{bennis@ee.oulu.fi}\\
$^{4}$Mathematical and Algorithmic Sciences Lab, Huawei France R\&D, Paris, France, Email: \protect\url{merouane.debbah@huawei.com}\\
}\vspace*{-1em}
    \thanks{This research was supported by the U.S. National Science Foundation under Grants CNS-1460316, CNS-1526844, CNS-1836802, and CNS-1617896, Academy of Finland project
    (CARMA), and Thule Institute strategic project (SAFARI).}%
  }

%
%
%
%
\maketitle
\vspace{-0em}
\begin{abstract}
Next-generation wireless networks must enable emerging technologies such as augmented reality and connected autonomous vehicles via wide range of wireless services that span enhanced mobile broadband (eMBB), as well as ultra-reliable low-latency communication (URLLC). Existing wireless systems that solely rely on the scarce sub-6 GHz, microwave ($\mu$W) frequency bands will be unable to meet such stringent and mixed service requirements for future wireless services due to spectrum scarcity. Meanwhile, operating at high-frequency millimeter wave (mmWave)  bands is seen as an attractive solution, primarily due to the bandwidth availability and possibility of large-scale multi-antenna communication. However, even though leveraging the large bandwidth at mmWave frequencies can potentially boost the wireless capacity for eMBB services and reduce the transmission delay for low-latency applications, mmWave communication is inherently unreliable due to its susceptibility to blockage, high path loss, and channel uncertainty.  
	Hence, to provide URLLC and high-speed wireless access, it is desirable to seamlessly integrate the reliability of $\mu$W networks with the high capacity of mmWave networks. To this end, in this paper, the first comprehensive tutorial for \emph{integrated mmWave-$\mu$W} communications is introduced. This envisioned integrated design will enable wireless networks to achieve URLLC along with eMBB by leveraging the best of two worlds: reliable, long-range communications at the $\mu$W bands and directional high-speed communications at the mmWave frequencies. To achieve this goal, key solution concepts are discussed that include new architectures for the radio interface, URLLC-aware frame structure and resource allocation methods along with mobility management, to realize the potential of integrated mmWave-$\mu$W communications. The opportunities and challenges of each proposed scheme are discussed and key results are presented to show the merits of the developed integrated mmWave-$\mu$W framework.

\end{abstract}

\section{Introduction} \vspace{-0cm}
 \subsection{Motivation and Problem Statement}
Next-generation wireless networks must support a new breed of heterogeneous wireless services, ranging from ultra-reliable low-latency traffic to bandwidth-intensive applications. To date, \emph{ultra-reliable low-latency communication (URLLC)} has been typically viewed as a solution for applications with only short packets (e.g., factory automation) that do not require high data rates and do not share the radio spectrum with enhanced mobile broadband (eMBB) services. \emph{However, with the advent of new applications, such as virtual reality (VR) and haptics \cite{MZVR}, wireless network designs must support URLLC jointly with eMBB communications.}  
\begin{figure}[t!]
	\centering
	\centerline{\includegraphics[width=\columnwidth]{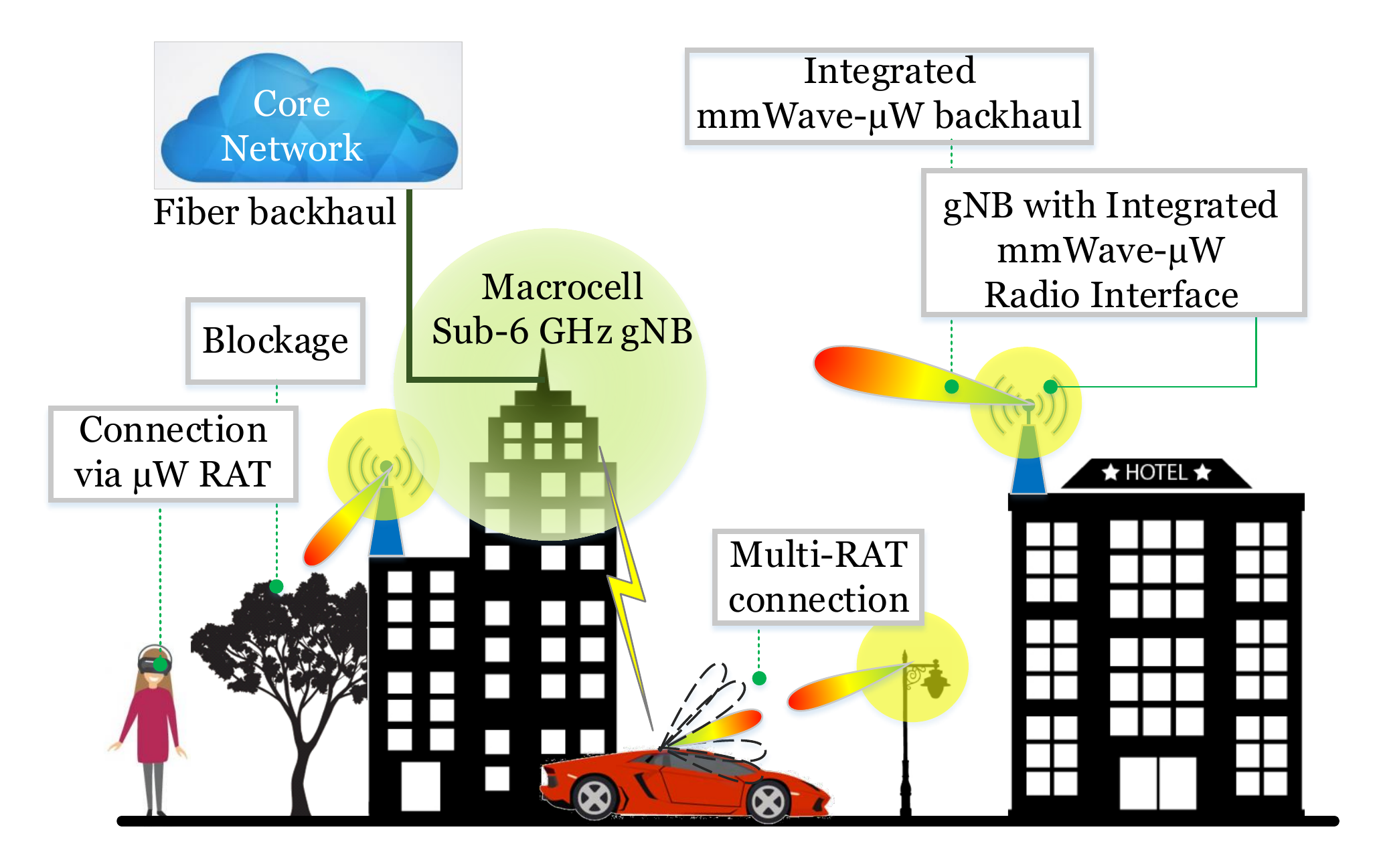}}\vspace{-0cm}
	\caption{\small Integrated mmWave-$\mu$W wireless network.}\vspace{-0.3cm}
	\label{fig1}\vspace{-.5em}
\end{figure}	

Two key technologies that will have to meet these stringent requirements include: 1) VR and augmented reality (AR) and 2) Connected autonomous vehicles, as shown in Fig. \ref{fig1}. In VR/AR applications, one solution to reduce the size of wireless headsets is to offload computational tasks, such as video rendering, to an edge computing server. However, such online processing requires transmission of large data packets over the wireless channel (in real-time) with high reliability and ultra low latency. In fact, the so-called ``motion-to-photon'' latency in VR/AR applications must be less than 20 ms, otherwise, it is not possible to guarantee an immersive virtual experience for the user. In addition, connected autonomous vehicles also rely on a mixed set of URLLC and eMBB services. In fact, vehicles must exchange large uncompressed sensing information to build high definition (HD) maps in real-time. Meanwhile, other autonomous vehicle applications such as pre-crash sensing and see-through will impose latency requirements up to 20 ms and 50 ms, respectively.



While cellular communications, including the recent specifications in 3rd Generation Partnership Project (3GPP) release 15 (Rel-15), are mainly optimized to increase coverage and data rate for eMBB, achieving URLLC together with eMBB for these applications requires a major rethinking of how cellular networks must be designed and operated.

\subsection{Background and Challenges}\label{secII}
In June 2018, the standalone mode for the fifth generation (5G) network has been finalized
	in 3GPP Rel-15 which enables 5G systems -- new radio (NR) and core network -- to operate independent from
	 existing cellular systems \cite{3GPPTS38300}. 5G networks are anticipated to support two frequency ranges, namely \emph{sub-6 GHz, microwave ($\mu$W)} (frequency
	range 1) and \emph{millimeter wave (mmWave)} (frequency range 2), each composed of several frequency
	bands. Due to the promising spectrum
	availability, mmWave communication is viewed as the key enabler for eMBB applications. Operating at mmWave frequency bands
	also allows the implementation of small-sized antenna arrays
	with large number of elements that can facilitate pencil-beam
	directional transmissions and extend the transmission range. This is particularly attractive for small form factor devices, such as wearable VR/AR equipments, that require both URLLC and high data rates. 
	 In addition, next-generation NodeBs (gNBs) that operate at mmWave frequencies do not interfere with $\mu$W gNBs. Hence, mmWave gNBs can be deployed within the coverage areas of $\mu$W gNBs in order to remove coverage holes, offload traffic, provide high data rates, and reduce the over-the-air transmission latency. Owing to these unique characteristics, mmWave communication is also foreseen as an enabler for fast wireless backhaul connectivity, particularly in ultra dense network deployments.

Nonetheless, mmWave signals cannot easily penetrate obstacles, and consequently, mmWave links
	are highly intermittent \cite{Rangan14}. This will be a real challenge for mobile applications, such as connected autonomous vehicles or VR/AR, since a mobile user may frequently experience blockage caused by buildings, vehicles,
	vegetation, humans, or urban furniture. Consequently, it is difficult to guarantee high reliability at the
	mmWave frequency range. 

Despite the substantial work in the literature, prior art \textcolor{black}{(e.g., puncturing or superposition schemes discussed by 3GPP for multiplexing URLLC and eMBB services)} mainly aims to realize URLLC solely at the sub-6 GHz frequency range, due to the link layer challenges of mmWave communications. While such $\mu$W-centric solutions may yield decent coverage and provide URLLC for \textcolor{black}{intermittent} short packet transmissions, they cannot support applications that require eMBB \textcolor{black}{together with real-time} URLLC traffic, such as VR/AR and connected autonomous vehicles. Furthermore, existing solutions for addressing mmWave link-layer challenges, such as blockage and beam training, also require substantial control overhead among mmWave gNBs and users \cite{7959169,7876982}. As such, 
there is an academic \cite{8107708,7959177,review2,review3,mehdi2018,review1,7929424} and industrial \cite{white3,white1} consensus with regard to the importance of \emph{mmWave-$\mu$W radio access technology (RAT) integration} as a  cost-effective solution to achieve high capacity, reliability, and low latency for emerging wireless services.
\subsection{Objectives and Key Contributions}
The main contribution of this article is a comprehensive roadmap towards realizing  the vision of \emph{integrated mmWave-$\mu$W} wireless networks for enabling cellular systems that jointly support URLLC and eMBB services.   To this end, we first introduce a new radio interface to enable the integration of mmWave and $\mu$W communications in a cellular network. Building on the integrated radio interface, we discuss a new frame structure design with flexible numerology and transmission time interval (TTI) to enable dynamic scheduling of URLLC and eMBB traffic jointly over mmWave and $\mu$W frequencies. To study the performance of this framework, we explain the key factors that impact the E2E performance, while taking into account the introduced integrated radio interface along with the flexible radio access network (RAN) architecture in 5G NR. Finally, we introduce novel schemes that leverage the developed integrated mmWave-$\mu$W framework to efficiently manage heterogeneous traffic over the mmWave and $\mu$W frequency bands. Together, the discussed solutions can potentially reap the full benefits of millimeter wave and sub-6 GHz communications thus providing unique opportunities to enable URLLC jointly with eMBB for wireless access and backhaul connections.  


 \section{Integrated mmWave-$\mu$W Radio Interface}\label{Sec:II}
\begin{figure}[t!]
	\centering
	\centerline{\includegraphics[width=\columnwidth]{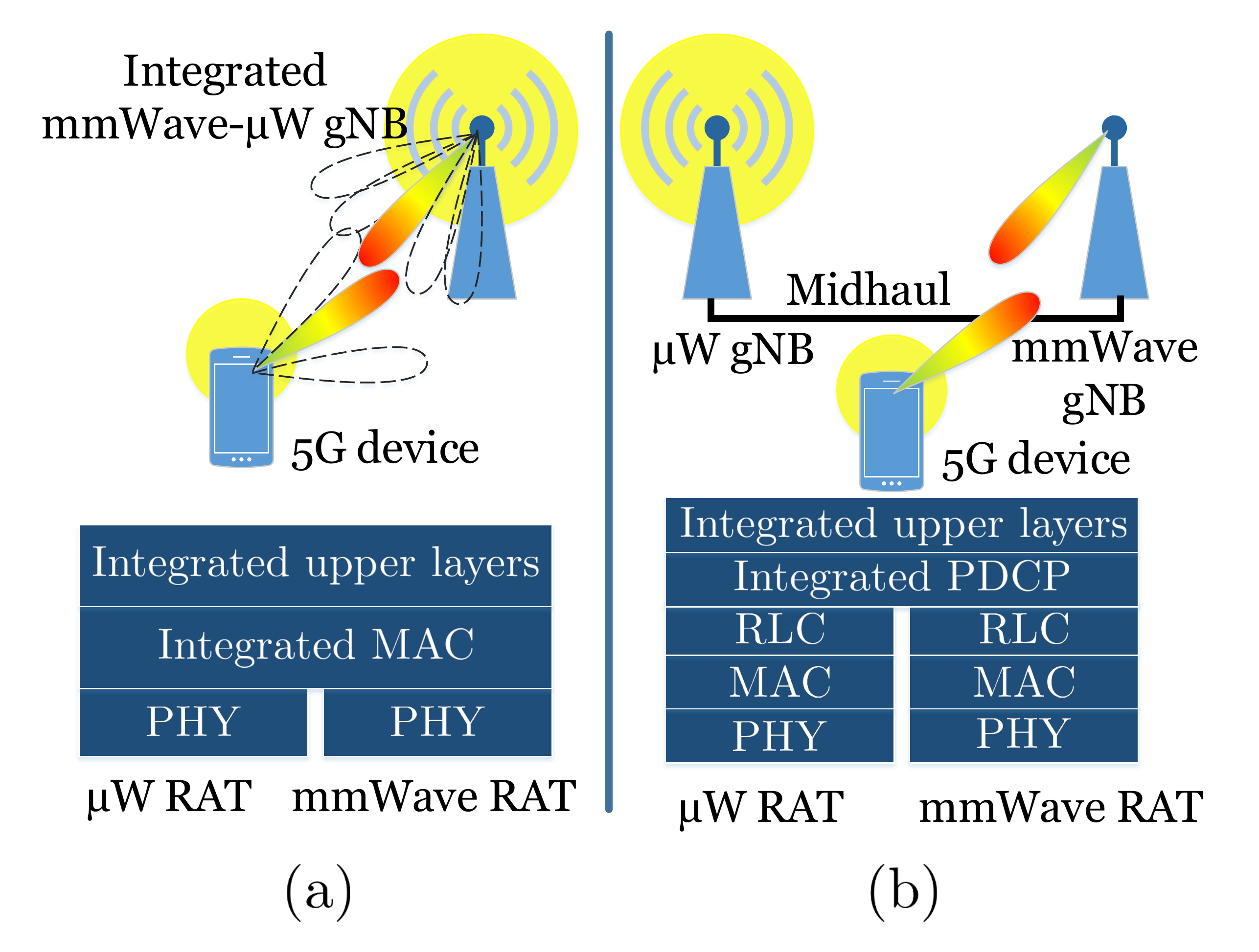}}\vspace{0cm}
	\caption{\small Integration schemes: (a) MAC layer integration, (b) PDCP layer integration.}
	\label{integration}\vspace{-1em}
\end{figure}

\vspace{-0em}

One of the first steps towards supporting both URLLC and eMBB traffic for a given user is to enable simultaneous user-planes at mmWave and sub-6 GHz frequency ranges. Nonetheless, the existing radio interface protocol stack in 5G NR only supports user-plane at a single frequency range. Hence, we explore the opportunities for integration schemes at medium access control (MAC) and packet data convergence protocol (PDCP) layers, shown in Fig. \ref{integration}.

Fig. \ref{integration}(a) presents the integration at the MAC layer of the radio interface. This scheme facilitates \emph{inter-frequency range carrier aggregation}, enabling a user to simultaneously receive or transmit on one or multiple component carriers at  frequency range 1 or 2. As shown in Fig. \ref{integration}(a), this scheme
	allows both mmWave and sub-6 GHz radios to be located at the same cell site, thus eliminating the need
	for a \emph{midhaul} connection between the two radio interfaces. This tight integration, in turn, enables fast scheduling of the traffic
	at both mmWave and sub-6 GHz. Such capability is critical for joint management of URLLC and eMBB traffic.

To fully exploit the MAC layer integration in NR, the following
challenges must be addressed. First, the frame structure considered in NR has a fixed one millisecond (ms) subframe duration, allowing different numerology at each subframe. With such a large subframe duration, it is not possible to achieve robustness against blockage at the mmWave band. Additionally, NR numerology options
are designed for scenarios with a single frequency range. With the MAC layer integration,
more numerology options are needed for joint scheduling of the traffic at the aggregated mmWave and sub-6
GHz frequencies. 

Meanwhile, the scheme in Fig. \ref{integration}(b) offers more flexible deployments, since mmWave and $\mu$W RATs can belong to different gNBs. This scheme provides an integration at the  PDCP layer and is suitable for ultra-reliable communications, as it enables a user to 
simultaneously connect to multiple gNBs. That is, if the connection over the mmWave RAT is interrupted, packets can be forwarded to users over the $\mu$W RAT. PDCP layer integration also facilitates \emph{packet duplication} by transmitting the same data over both mmWave and $\mu$W RATs thus achieving transmission diversity gains to maximize the reliability for URLLC traffic. One shortcoming of this scheme is the need for having a high capacity midhaul link between the mmWave and $\mu$W gNBs. In addition, packet forwarding over the midhaul link can result in an excessive delay which may not be tolerated by low-latency applications.

To exploit the desirable features of each integration scheme in Fig. \ref{integration}, we advocate a hybrid
architecture which incorporates both MAC and PDCP
layer integration.
Such a hybrid architecture will also allow for backward
compatibility with legacy single-RAT cellular
networks. Here, we note that other integration
schemes -- which mostly will lie between
the two proposed schemes in Fig. \ref{integration} -- can also
be considered. For example, the fast session
transfer (FST) introduced in IEEE 802.11ad
WLAN protocol can be considered as an integration scheme with a separate MAC and PHY layers at mmWave and $\mu$W RATs. However, these approaches are not as efficient as the MAC layer integration in terms of latency,
since they require substantial control overhead (including a negotiation phase) to switch between RATs.

\section{Frame Structure and RAN Architecture for Integrated mmWave-$\mu$W Networks}\label{sectionIII}
   
To benefit from the proposed integration schemes, new designs for frame structure and RAN architecture are required so as to deliver seamless multiplexing of URLLC and eMBB traffic over mmWave and $\mu$W frequencies. Next, we first propose a new frame structure and then discuss the RAN architecture with mmWave and sub-6 GHz RAT integration. 

\subsection{Frame Structure}\label{IIIA}
To design the frame structure, we focus on the orthogonal frequency-division multiplexing (OFDM) waveform, as recommended by NR specifications. To determine the numerology (i.e., subcarrier spacing and
cyclic prefix length) for OFDM symbols, we must account for substantially different channel characteristics at mmWave and sub-6 GHz bands. These metrics are determined based on the power delay profile of the wireless channel at each frequency band. As a next step, we find two important parameters: 1) TTI as the duration of a schedulable transmission period on the radio link and 2) Subframe duration as the shortest time that the numerology at one frequency range can be updated, cognizant
of the channel state information at the other frequency
range. While these two parameters are conventionally
considered to be the same (e.g., in LTE systems), the
proposed frame structure will distinguish them in order
to include dynamic numerology selection.

This flexible frame structure will enable multiplexing of
eMBB and URLLC services, while yielding robustness against the uncertainties of the
mmWave link quality and coping with the large traffic at the sub-6 GHz
frequency range. For example, if the mmWave link becomes
unavailable (due to blockage or beam misalignment),
delay-intolerant services that are already scheduled
over the mmWave resources must be re-scheduled
at the sub-6 GHz frequency band. The proposed dynamic
numerology and TTI selection will allow the sub-6 GHz band to accommodate this traffic that was moved from the
mmWave frequency range.

Denoting the TTI by $\tau$, it can be chosen as $\tau=2^i t_s$, where $i$ takes nonnegative integer values and $t_s$ is
the OFDM symbol duration. Furthermore, the $1$ ms subframe duration that is considered in NR is not short
enough for such dynamic traffic scheduling in an integrated mmWave-$\mu$W radio interface. That is because
mmWave channel variations can be much faster, due to their large Doppler spread.
Coupled with the mobility of users (such as connected autonomous vehicles with high speeds), the NR subframe duration must be revised. Consequently, we consider three link states at the mmWave radio interface: 1) Line-of-Sight (LoS),
2) Non-LoS (NLoS), and 3) Blocked. With a LoS mmWave link state, bandwidth-intensive applications, such
as HD-maps or 3D VR video streaming, can be scheduled at the mmWave frequency band. Thus, we
can fully dedicate the sub-6 GHz band to URLLC services by choosing very small TTIs (e.g., $\tau= t_s$). In
case of NLoS or blocked mmWave link states, we can update the numerology and TTI at the sub-6 GHz
radio interface to allow time or frequency multiplexing of the migrated services from the mmWave to sub-6
GHz band. Using the proposed flexible frame structure and considering the mmWave link states, suitable numerology, TTI, and subframe combinations in terms of the user-plane
latency, reliability, and data rate, can be determined. 

\begin{figure}[t!]
	\centering
	\centerline{\includegraphics[width=\columnwidth]{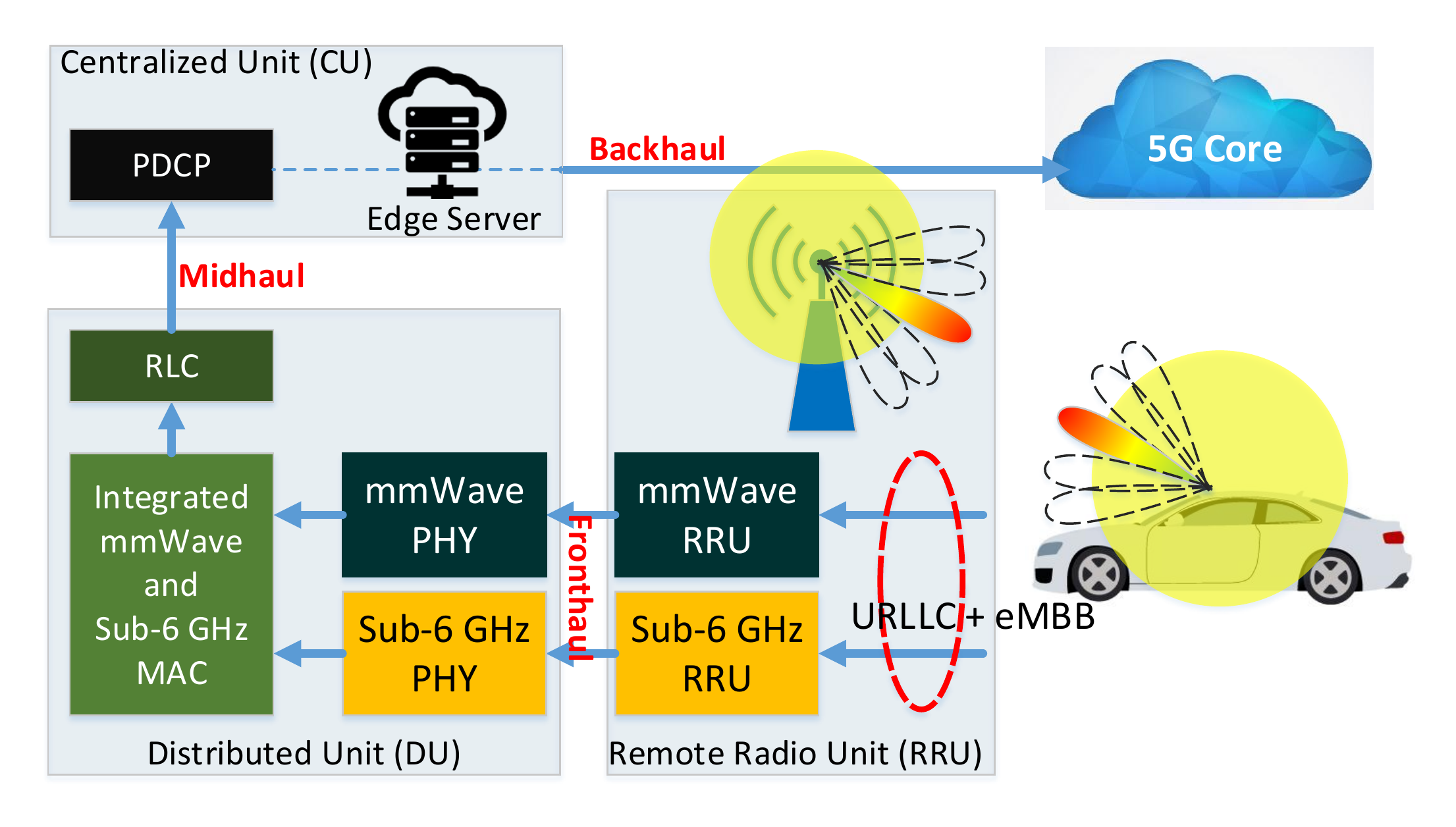}}\vspace{-.2cm}
	\caption{\small Uplink user-plane in the proposed NG-RAN with MAC-layer integration.}
	\label{RAN}\vspace{-1em}
\end{figure}
\subsection{Key E2E Performance Indicators}
The proposed
	frame structure will facilitate dynamic traffic management over the aggregated mmWave and
	sub-6 GHz frequency bands. Supported by this flexible frame structure, our next goal is to analyze the E2E performance of the next-generation RAN (NG-RAN) that supports the proposed MAC-layer and PDCP-layer integration schemes. Here, we note that existing design paradigms for cellular communications mostly rely on optimizing the network's average performance in terms of spectral efficiency, coverage, and outage, among other metrics. Nonetheless, depending on the use case, URLLC requires meeting \emph{instantaneous}  -- per packet -- latency and reliability constraints. To further clarify the new URLLC-centric quality-of-service (QoS) requirements, next, we overview the 3GPP definitions for latency and reliability. 
	
	\textbf{E2E latency}: According to the 3GPP \cite{3gpp3}, E2E latency is defined as the time it takes to transfer a given piece of information from a source to a destination, measured at the communication interface, from the moment that it is transmitted by the source to the moment it is successfully received at the destination.

	\textbf{Reliability}:  3GPP \cite{3gpp3} defines reliability as the percentage value of the amount of sent network layer packets successfully delivered to a given node within the time constraint required by the targeted service, divided by the total number of sent network layer packets.

	 To make these generic definitions specific to the integrated mmWave-$\mu$W wireless networks, we note that 5G NG-RAN has a flexible architecture with different RAN splitting option. These splitting options divide the gNB's radio
	protocol stack within three sub-systems: 1) Remote radio units (RRUs), 2) Distributed unit (DU), and 3)
	Centralized unit (CU). Accordingly, the NG-RAN will include the fronthaul (link between
	RRU and DU), midhaul (link between DU and CU), and backhaul links (link between CU and the
	core network), depending on the splitting option. Fig. \ref{RAN} shows the uplink user-plane in the proposed
	NG-RAN structure with MAC-layer integration. This scheme will realize separate Layer 1 functions and shared Layer 2 and 3 functions
	for the mmWave and sub-6 GHz radio access technologies. As mentioned earlier, the key difference between this architecture  and  the PDCP-layer integration, is its ability to realize a single MAC function for managing both mmWave and sub-6 GHz radio interfaces.

As such, we can clearly define the key E2E performance indicators for the NG-RAN. For example, the one-way E2E latency for the proposed NG-RAN in Fig. \ref{RAN} comprises the over-the-air transmission delay (at wireless access, fronthaul, and midhaul), as well as the transceiver's queuing delay. Additionally, if the requested wireless service requires processing at an edge server (can be located within the CU of a gNB), then the task completion time must also be taken into account. 

Building on these performance metrics, we note that the regime for defining \emph{high} reliability or \emph{low} latency depends on the specific use case scenario. For example, as  discussed in Section I, the latency required for autonomous vehicle services can vary from less than ten milliseconds, up to 50 ms. Moreover, reliability constraint can vary from $ 99.9 \%$ for monitoring applications to larger values for remote control  \cite{3gpp3}. Therefore, potential solutions for realizing URLLC are application specific. Here, we focus on the previously mentioned wireless services whose URLLC requirements are not restricted to those required by conventional short-packet applications.

\section{Resource and Mobility Management in Integrated mmWave-$\mu$W Wireless Networks}
\vspace{.5em}
\subsection{Multiple Access and Resource Management}
According to the 3GPP \cite{3gpp3}, one of the key elements needed to achieve the target performance in next-generation wireless networks is to develop fast, scalable, and efficient resource management algorithms. Particularly, in URLLC with relatively shorter TTIs, it is essential to reduce the signaling overhead at the control-plane for user scheduling.  Achieving this goal at the mmWave frequencies is very challenging, mainly due to the directionality of the mmWave links and their susceptibility to blockage.   Most of the existing mmWave protocols employ a time division multiple access (TDMA) scheme for resource management, primarily due to the lower complexity for beam-training with only one user at each time slot. However, TDMA is not suitable to support low latency services, particularly if the number of users is large within the coverage area. Additionally, user multiplexing (spatially or at different frequencies) requires substantial overhead to perform channel estimation, which again affects the time for establishment of a mmWave link.

Considering the aforementioned challenges, the integration of mmWave and $\mu$W RATs will provide the following opportunities for resource management: 1) Minimizing overhead and delay for beam-training at the mmWave RAT by using out-of-band   channel measurements and localization at the $\mu$W frequencies, 2) High reliability for access and backhaul links by switching to the $\mu$W RAT and retaining connection, once the mmWave transmission is interrupted by blockage, 3) Control-plane and user-plane signals  can be handled over different RATs, 4) Enabling fast and seamless RAT switching, depending on the link quality, QoS requirements, and the load at each RAT, 5) Reducing the number of link hops in wireless backhaul networks by leveraging $\mu$W links to penetrate obstacles, if needed, and 6) Managing uplink and downlink traffic over different RATs.

Despite these unique opportunities, resource management and user-cell association in integrated mmWave-$\mu$W systems is more challenging and complex compared with conventional single RAT networks. Such higher complexity mainly stems from having completely different channel characteristics at the mmWave and $\mu$W frequencies. For instance, in resource allocation, fast fading and shadowing can result in frequent RAT switching and significant signaling overhead at the control-plane. Furthermore,  traditional user-cell association schemes, based on the maximum received signal strength indicator (max-RSSI) or maximum signal-to-interference-plus-noise ratio (max-SINR) will result in substantially unbalanced load distribution across mmWave and $\mu$W RATs. That is because the path loss at mmWave frequencies is larger than $\mu$W bands, thus, the max-RSSI scheme will associate most of users to the $\mu$W RAT. In addition, given that interference at mmWave frequencies is less severe than at the $\mu$W bands (due to link directionality, as well as poor penetration and scattering of mmWave signals),  the max-SINR scheme will assign more users to the mmWave RAT. Such unbalanced load distributions can introduce substantial latency at the MAC layer for link establishment and scheduling of the users.

 \begin{table*}[t!]
 	\scriptsize
 	\centering
 	\caption{
 		\vspace*{-0em}\small Summary of proposed frameworks to realize integrated mmWave-$\mu$W wireless networks.}\vspace*{-1em}
 	\renewcommand{\arraystretch}{1.5}
 	
 	\begin{tabular}{| >{\centering\arraybackslash}m{1.5cm}| >{\centering\arraybackslash}m{5cm}|>{\centering\arraybackslash}m{5cm}|>{\centering\arraybackslash}m{4.5cm}|}
 		\hline
 		\bf{Proposed framework}	&\bf{Integrated mmWave-$\mu$W radio interface and frame structure} &\bf{Multiple access and resource allocation in integrated mmWave-$\mu$W networks} & \bf{\centering{Mobility management in integrated mmWave-$\mu$W networks}}\\
 		\hline
 		\hline
 		\multirow{6}{*}{\bf{Opportunities}}	& Fast scheduling of URLLC and eMBB traffic over the integrated mmWave and sub-6 GHz frequencies via MAC layer integration.& Minimizing overhead and delay for beam-training at the mmWave RAT by using out-of-band   channel measurements and localization at the $\mu$W frequencies. & \multirow{-1.5}{4cm}{\centering Facilitating mmWave communications for mobile users with high speeds.}\\
 		\cline{2-4}
 		& Facilitating multi-connectivity and packet duplication via PDCP-layer integration to maximize the reliability. & Control-plane and user-plane separation across frequency ranges 1 and 2, supported by the proposed integrated radio interface. & Increasing reliability by reducing handover failures. \\
 		\cline{2-4}
 		&  \multirow{-.6}{4.5cm}{\centering Flexible frame structure with joint mmWave and $\mu$W dynamic numerology design to cope with the intermittent nature of mmWave links, increase reliability, and reduce latency.} &  Joint uplink-downlink traffic management and load balancing over the integrated mmWave-$\mu$W radio interface. &   \multirow{-1.5}{4.5cm}{\centering Increasing reliability by managing control signals over the $\mu$W bands, while offloading data traffic to mmWave frequencies.}\\
 		\cline{3-4}
 		& & Fast and reliable backhaul connectivity, particularly in multi-hop backhaul networks. & Enabling data caching via the mmWave RAT to avoid frequent handovers, while traversing small cells.\\
 		\hline
 		\hline
 		\multirow{4}{*}{\bf{Challenges}}	&Higher complexity for MAC protocols, compared with single-RAT networks.	& Substantially different channel characteristics over the mmWave and $\mu$W frequencies. & Higher complexity for mobility management, given more options for vertical and horizontal handovers.	 \\
 		\cline{2-4}
 		&Designing application-centric parameters for the frame structure. &User-cell association and load balancing is complicated and challenging. &{\centering Deriving theoretical bounds for latency and reliability in mobile scenarios.} \\
 		\cline{2-4}
 		& \multicolumn{3}{c|}{\centering E2E performance analysis of integrated mmWave-$\mu$W networks.}\\
 		\hline
 	\end{tabular}\label{tab1}\vspace{-1.5em}
 \end{table*}

\vspace{-.5em}
\subsection{Mobility Management} 
User mobility will exacerbate the challenges of URLLC, due to the following reasons: \emph{First}, mobile users, such as autonomous vehicles, will experience frequent handovers (HOs), while passing through small cells, which naturally increases the overhead and communication latency. Such frequent HOs will also degrade reliability by increasing handover failure (HOF), particularly for users that are moving at high speeds. In fact, due to the small and disparate cell sizes in emerging cellular networks, users will not be able to successfully finish the HO process by the time they trigger HO and pass a target small cell. \emph{Second}, the inter-frequency measurements needed to  discover target small cells can be excessively power consuming and detrimental for the battery life of users, especially in dense networks with frequent HOs. \emph{Third}, guaranteeing a bounded latency for mobile applications is still a challenging open problem, especially with the imminent deployment of highly-mobile users such as drones or connected autonomous vehicles. That is because $\mu$W frequencies are often congested, and thus, the unmanageable traffic resulting from frequent HOs by mobile users will introduce unacceptable queuing delays. Thus, solely relying on $\mu$W frequencies to address mobility management can result in inefficient resource utilization and also limits the available frequency resources for the static users.

In this regard, the proposed mmWave-$\mu$W integration scheme in Section \ref{Sec:II} can provide new opportunities to address the aforementioned latency and reliability challenges.  The key idea is disjointing the control-plane and user-plane traffics by leveraging high-speed mmWave links for data transmissions, while using the $\mu$W RAT mainly for handling control and paging information. This flexible architecture can be used to cache large data, such as HD-maps for autonomous vehicles, via available high-speed mmWave links. Using the cached content, the mobile user can mute HO with small cells and only maintain the $\mu$W connection to a macro-cell base station for communicating necessary control signals. Therefore, this scheme can alleviate the reliability issues associated with HOFs by avoiding unnecessary data connections and HOs with small cells. Moreover, the network will be able to offload the traffic of mobile users from heavily utilized $\mu$W frequencies to mmWave frequencies and substantially reduce the traffic and over-the-air latency at the $\mu$W network.

\begin{figure}[t!]
	\centering
	\centerline{\includegraphics[width=8cm]{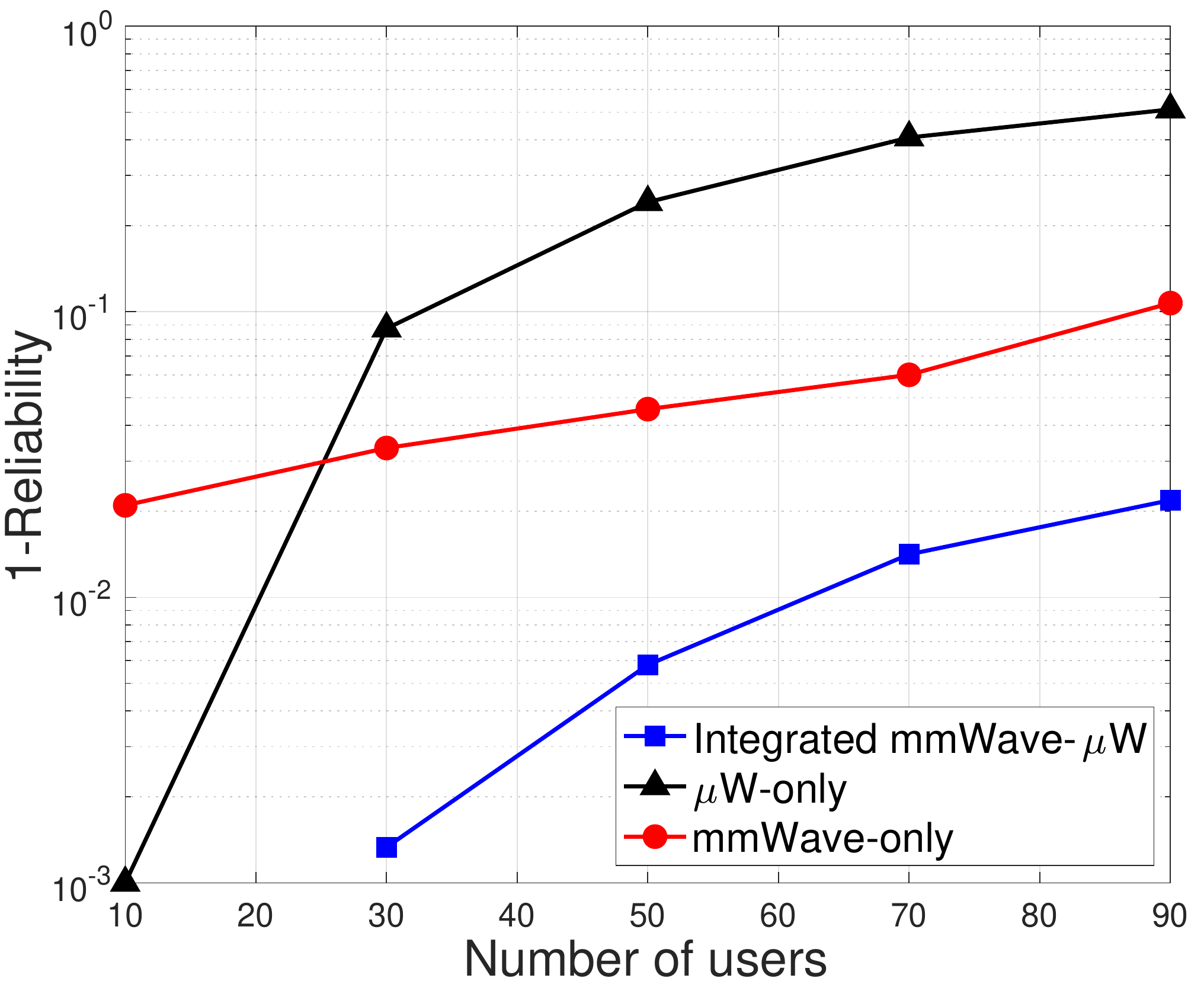}}\vspace{-.2cm}
	\caption{\small Reliability comparison between integrated mmWave-$\mu$W  networks and single-RAT mmWave and $\mu$W networks as function of network size.}\vspace{-1em}
	\label{sim1}
\end{figure}

\section{Performance of Integrated mmWave-$\mu$W Wireless Networks: Proof of Concept}
To demonstrate the performance gains that the proposed integrated mmWave-$\mu$W framework can achieve, we consider two scenarios: 1) Resource allocation in dense networks with delay-constraint applications and 2) Mobility management in mobile wireless networks.
\subsection{Resource Allocation in Integrated mmWave-$\mu$W Wireless Networks}

Consider a MAC layer integration of mmWave and $\mu$W RATs at a small base station with $30$ dBm transmit power. The TTI duration is $1$~ms and the packet size for the downlink traffic is $20$ kbits. The available bandwidth at the mmWave and $\mu$W bands is $1$ GHz and $10$ MHz, respectively. We consider two types of traffic with $10$ ms and $20$ ms latency requirement. Users are distributed uniformly within the cell coverage and each user is associated randomly with one traffic type.

Fig. \ref{sim1} shows a performance comparison between the proposed integrated mmWave-$\mu$W framework and single-RAT networks. Here, reliability is calculated using the definition in  Section \ref{secII}. The results in Fig. \ref{sim1} demonstrate that the mmWave  network is not reliable and cannot achieve an acceptable performance, even for a network with $10$ users. In addition, we can easily observe that the reliability of the $\mu$W network rapidly decreases, as the number of users increases. Thus, the $\mu$W network cannot support the latency requirements of users in dense scenarios. In contrast, the proposed integrated scheme substantially improves the reliability of the network by managing the traffic jointly at mmWave and $\mu$W RATs.

\subsection{Mobility Management in Integrated mmWave-$\mu$W Wireless Networks}
\begin{figure}[t!]
	\centering
	\centerline{\includegraphics[width=8cm]{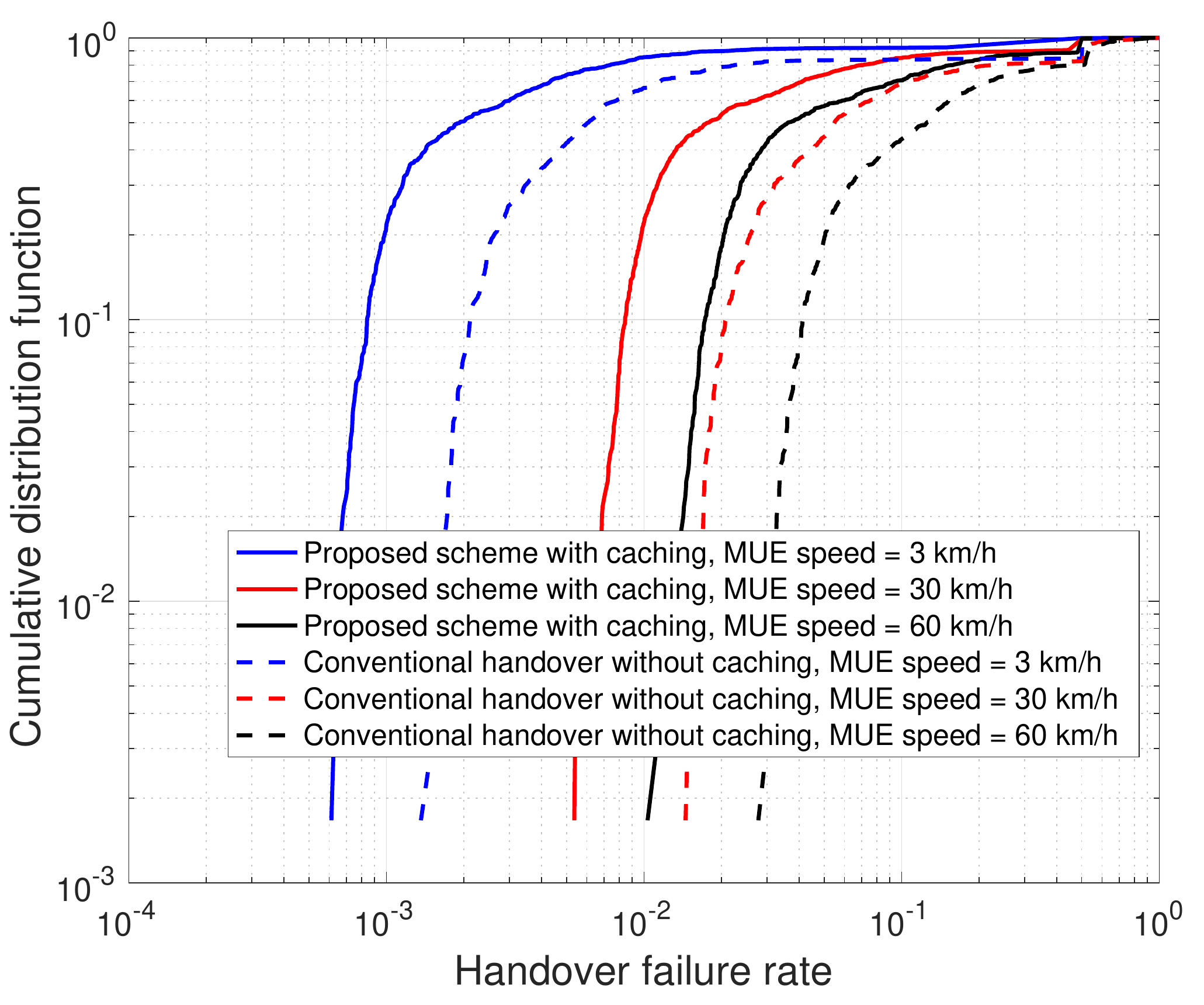}}\vspace{-.2cm}
	\caption{\small CDF of HOF vs. different user speeds.}\vspace{-1em}
	\label{sim2}
\end{figure}

While the previous example shows the merit of the proposed integrated framework in static networks, here, we consider a mobile scenario 
in which a set of mobile users (e.g. autonomous vehicles) move using random and fixed directions within a small cell network with $50$~base stations.	The mmWave RAT operates at the E-band with $5$ GHz bandwidth. Fig. \ref{sim2} shows the achievable performance gains of the mobility management scheme, proposed in Subsection \ref{IIIA}. Here, we developed a novel mobility management framework that allows users to mute their handover and cell search process, while traversing small cells. This scheme was realized based on the analytical methods proposed in \cite{8107708}, which leverages high-speed mmWave links, whenever available, to cache content at the mobile user device. Meanwhile, the control and paging information are handled over the $\mu$W frequencies. 
Fig. \ref{sim2} shows that the proposed framework can substantially improve reliability by decreasing the handover failure rate, even for users with high speeds.

\vspace{0em}
\section{Conclusions}
In this paper, we have provided a comprehensive
tutorial on the notion of integrated mmWave-$\mu$W wireless networks to enable emerging wireless services that require mobile broadband along with ultra-reliable low-latency communications. To achieve this network convergence, we have introduced new designs for the radio interface and frame structure to facilitate fast and reliable scheduling of eMBB and URLLC traffic jointly over mmWave and $\mu$W frequencies. Additionally, we have introduced new schemes for resource allocation and mobility management in integrated mmWave-$\mu$W wireless networks and discussed their opportunities and challenges. As summarized in Table \ref{tab1}, the proposed integrated mmWave-$\mu$W wireless network can reap the full benefit of multi-RAT capability to efficiently multiplex eMBB and URLLC services in next-generation wireless networks, achieve ultra-reliable communications capable of coping with the intermittent nature of mmWave links, and reduce handover failure rate in mobile networks.  Therefore, the proposed schemes  will pave the way for a seamless integration of mmWave technology into future wireless networks to support URLLC along with mobile broadband for emerging wireless technologies.

\bibliographystyle{IEEEtran}
\bibliography{references,bib}

\end{document}